\documentclass[main]{sshArticle}

\newcommand{\tas}{\texorpdfstring{TaS\textsubscript{2}}{TaS2}}
\newcommand{\tase}{\texorpdfstring{TaSe\textsubscript{2}}{TaSe2}}
\newcommand{\tasse}{\texorpdfstring{TaS\textsubscript{x}Se\textsubscript{2-x}}{TaSxSe2-x}}

\newcommand{\comment}[1]{}
\mathchardef\mhyphen="2D

\title{Endotaxial Stabilization of 2D Charge Density Waves with Long-range Order}

\author[1]{Suk~Hyun~Sung}
\author[1]{Nishkarsh~Agarwal}
\author[2]{Ismail~El~Baggari}
\author[3]{Yin~Min~Goh}
\author[4]{Patrick~Kezer}
\author[5]{Noah~Schnitzer}
\author[6]{\\Yu~Liu}
\author[6]{Wenjian~Lu}
\author[6,7,8]{Yuping~Sun}
\author[9,10]{Lena~F.~Kourkoutis}
\author[1,11]{John~T.~Heron}
\author[12]{Kai~Sun}
\author[1,11,*]{Robert~Hovden}

\affil[1]{Department of Materials Science and Engineering, University of Michigan, Ann Arbor, MI 48109}
\affil[2]{Rowland Institute at Harvard University, Cambridge, MA 02142}
\affil[3]{John A. Paulson School of Engineering and Applied Sciences, Harvard University, Cambridge, MA 02138}
\affil[4]{Department of Electrical and Computer Engineering, University of Michigan, Ann Arbor, MI 48109}
\affil[5]{Department of Materials Science and Engineering, Cornell University, Ithaca, NY 14853}
\affil[6]{Key Laboratory of Materials Physics, Institute of Solid State Physics, Chinese Academy of Sciences, Hefei 230031, People’s Republic of China}
\affil[7]{High Magnetic Field Laboratory, Chinese Academy of Sciences, Hefei 230031, People's Republic of China}
\affil[8]{Collaborative Innovation Centre of Advanced Microstructures, Nanjing University, 210093, Nanjing, People’s Republic of China}
\affil[9]{School of Applied and Engineering Physics, Cornell University, Ithaca, NY 14853}
\affil[10]{Kavli Institute at Cornell for Nanoscale Science, Ithaca, NY 14853}
\affil[11]{Applied Physics Program, University of Michigan, Ann Arbor, MI 48109}
\affil[12]{Department of Physics, University of Michigan, Ann Arbor, MI 48109}
\affil[*]{e-mail: hovden@umich.edu}
\date{\today}

\firstAuthorLastName{Sung}

\graphicspath{ {./Figs/} }

\abstract{

Charge density waves are emergent quantum states that spontaneously reduce crystal symmetry, drive metal-insulator transitions, and precede superconductivity. In low-dimensions, distinct quantum states arise, however, thermal fluctuations and external disorder destroy long-range order. Here we stabilize ordered two-dimensional (2D) charge density waves through endotaxial synthesis of confined monolayers of 1T-\tas{}. Specifically, an ordered incommensurate charge density wave (oIC-CDW) is realized in 2D with dramatically enhanced amplitude and resistivity. By enhancing CDW order, the hexatic nature of charge density waves becomes observable. Upon heating via in-situ TEM, the CDW continuously melts in a reversible hexatic process wherein topological defects form in the charge density wave. From these results, new regimes of the CDW phase diagram for 1T-\tas{} are derived and consistent with the predicted emergence of vestigial quantum order.

}

\begin{document}
\maketitle

Some exotic crystals spontaneously reorganize their valence electrons into periodic structures known as charge density waves (CDWs). In essence, two crystals emerge---the underlying atomic lattice and the emergent charge lattice. Just like atomic crystals, a charge density wave has defects: dislocations, disclinations, and elastic deformation~\cite{Savitzky_2017,ElBaggari_2018,Joe_2014}. Furthermore, the charge density wave can undergo phase transitions wherein the charge lattice unit cell changes shape and size. All of this CDW reshaping and topological restructuring occurs even when the underlying atomic lattice remains unchanged.

In low dimensions, these quantum phase transitions are promising candidates for novel devices~\cite{Hollander_2015, Liu_2016, Hsieh_2017, Tokura_2017}, efficient ultrafast non-volatile switching~\cite{Tsen_2015, Vaskivskyi_2016, Hellmann_2012}, and suggest elusive chiral superconductivity~\cite{Ganesh_2014, Kanigel_2020, Navarro_2016}. Unfortunately, 2D CDWs are inherently unstable and accessing low-dimensional CDWs remains a challenge~\cite{MerminWagner_1966, Hohenberg_1967, Yu_2015}. Even worse, at elevated temperatures where devices typically operate, disruption of charge density waves is all but guaranteed due to ever-present disorder~\cite{Imry_1975,Nie_2014,Yao_2022}. A long-range ordered incommensurate CDW has yet to be reported.

Here we stabilize ordered incommensurate charge density waves (oIC-CDW) at elevated temperatures (T\textsubscript{IC} = 350~K) in two-dimensions by endotaxial synthesis of \tas{} polytype heterostructures. The estimated hundred-fold amplitude enhancement of the charge density wave has an increased coherence length comparable to the underlying atomic crystal. The enhanced order of the oIC-CDW increases electronic resistivity. This substantial enhancement of charge order is achieved through encapsulation of an isolated octahedral \tas{} CDW layer within a matrix of prismatic \tas{} metallic layers via 2D endotaxial synthesis.

Realizing the ordered incommensurate CDW reveals CDWs have hexatic structure at high-temperature---that is, long-range translational symmetry is limited by proliferation of topological defects (i.e., dislocations and disclinations) in CDWs. We show at high-temperatures, the CDWs in \tas{} continuously melt as additional dislocations and disclinations form in the charge lattice.  This hexatic CDW melting process was not previously observable since the incommensurate CDW normally emerges as a highly-disordered, melted state. By restoring order through 2D endotaxy, we can reversibly melt and unmelt CDWs in \tas{}. Based on these results, we access new regimes of the CDW phase diagram for octahedrally coordinated \tas{} in temperature vs disorder space. Similar vestigial ordering (i.e., hexaticity) was predicted by Nie, Tarjus and Kivelson~\cite{Nie_2014}; however, with 2D endotaxy we can now tune down the disorder in the CDW phase diagram.

\begin{figure*}[ht]
    \centering
    \includegraphics[width=1.0\linewidth]{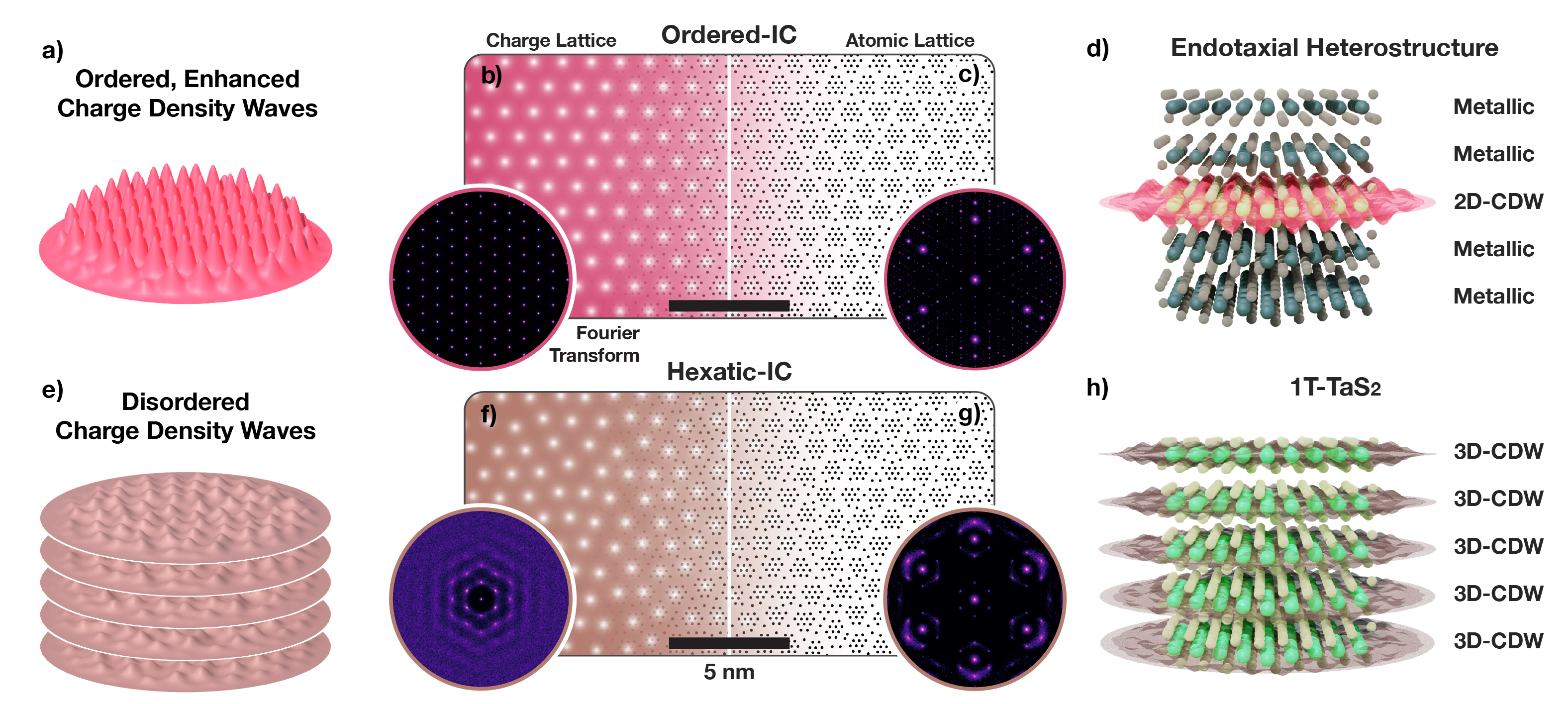}
    \caption{\textbf{Long-range Ordered Incommensurate Charge Density Waves.} a) Schematic representation of ordered IC-CDW. The CDW is two-dimensional with little disorder. b) Ordered IC-CDW illustrated as a crystalline charge-density lattice. Inset) Fourier transform of the charge lattice shows well defined peaks. c) Associated periodic lattice distortions (PLDs) move tantalum nuclei along the charge density gradient. Inset) Simulated diffraction shows sharp superlattice peaks decorating Bragg peaks. d) Schematic representation of ordered IC-CDW in endotaxial polytype heterostructure. Mono- or few layers of endotaxially protected Oc-\tas{} hosts 2D ordered IC-CDWs. e) Schematic representation of hexatic IC-CDW. The CDW phase is quasi-2D with non-trivial interlayer interactions, and hexatically disordered. f) Charge density distribution is comparable to hexatically disordered crystal lattice. Inset) Structure factor reveals azimuthally diffused peaks---characteristics of hexatic phases. g) Associated lattice distortion of IC-CDW with (inset) Fourier transform showing azimuthally blurred superlattice peaks while maintaining sharp Bragg peaks. h) Schematic representation of hexatic IC-CDW in bulk 1T-TaS$2$ where every layer hosts disordered IC-CDW.}
    \label{fig::Overview}
\end{figure*}

\section*{The Ordered Incommensurate Charge Density Wave}

The ordered incommensurate CDW (oIC) reported herein (Fig.~\ref{fig::Overview}a--d) is strikingly distinct from the well-known incommensurate (IC) CDW (Fig.~\ref{fig::Overview}e--h) found in 1T-\tas{} or 1T-\tase{}. Here, the oIC phase is a truly two-dimensional (2D) CDW with long-range positional and orientational order that couples strongly with the underlying crystal lattice (Fig.~\ref{fig::Overview}a). The oIC-CDW, illustrated in Figure~\ref{fig::Overview}b, is a crystalline charge-lattice with well-defined, sharp peaks in Fourier space (Fig.~\ref{fig::Overview}b-inset). This CDW charge-lattice (a\textsubscript{CDW} = 11.87~nm) exists within an underlying atomic lattice illustrated in Figure~\ref{fig::Overview}c.

Electron--lattice interaction is an essential aspect of CDWs, and associated soft-phonon modes manifest as static periodic lattice distortions (PLDs) that reduce crystal symmetry and lower the electronic energy~\cite{McMillan_1975, Peierls_1930}. For \tas{}, the CDW pulls atoms toward the nearest charge maximum to form periodic clusters of atoms (Fig.~\ref{fig::Overview}c). Notably for incommensurate charge ordering, each cluster is distinct since the atomic lattice is not commensurate with the CDW. While these lattice distortions are small ($<$10~pm), selected area electron diffraction (SAED) is sensitive to subtle picoscale distortions and making it a popular choice for characterization of CDW/PLDs ~\cite{Sung_2022_tPLD}. CDW/PLDs diffract incident swift electrons into distinct superlattice peaks decorating each Bragg peak~\cite{Wilson_1975, Overhauser_1971, Hovden_2016, ElBaggari_2018}. In reciprocal space, the CDW charge lattice (Fig.~\ref{fig::Overview}b-inset) and the measurable atomic superlattice peaks (Fig.~\ref{fig::Overview}c-inset) have corresponding spacing, symmetry, and intensity.

Diffracted superlattice peaks provide a direct measure of the CDW lattice and contain rich information on their order-disorder. Specifically, diffraction represents an ensemble average of the structure over the selected area, and disorder manifests as diffused diffraction peaks~\cite{Huang_1947,Peterson_1994}. Disorder of CDWs smears superlattice peaks but leaves the principle Bragg peaks unaffected~(Fig.~\ref{fig::Overview}g-inset). For oIC-CDWs, the charge lattice is ordered with limited defects, thus diffraction shows both sharp superlattice and Bragg peaks (Fig.~\ref{fig::Overview}c-inset). In contrast, the well-known IC-CDW in 1T-\tas{} possesses significant disorder of its charge distribution. Across decades, the IC phase in 1T-\tas{} is reported with a ring-like, azimuthally diffuse diffraction around each Bragg peak~\cite{Wilson_1975,Williams_1974, vanLanduyt_1974,Ishiguro_1991}, yet the origin of the diffused superlattice peaks is hardly discussed~\cite{Welberry_1987,Dai_1992}.

Here we present the well-known IC-CDW in bulk 1T-\tas{} as a hexatically disordered charge lattice containing dislocations and disclinations (Fig.~\ref{fig::Overview}f). In-situ SAED of 1T-\tas{} taken at 408~K (Fig.~\ref{fig::ThermalTreatment}a) shows azimuthally blurred first order superlattice peaks (marked brown). Averaging all six third order Bragg peaks (inset, $\Gamma_3$) better highlights this point. Notably, hexatic phases are known to have six-fold rotationally symmetric, azimuthally diffused peaks~\cite{Brock_1989}. The experimental diffraction of IC-CDWs are consistent with a hexatic charge distribution (Fig.~\ref{fig::Overview}f)~\cite{Kosterlitz_1973, Brock_1989, Dai_1992, Peterson_1994} and corresponding azimuthally diffuse structure factor (Fig.~\ref{fig::Overview}f, g-inset). The IC-CDWs are three-dimensional (or quasi-2D) with non-negligible out-of-plane interactions (Fig.~\ref{fig::Overview}e--h).

In contrast, the oIC-CDW, shows drastically sharper and stronger superlattice peaks measured by in-situ SAED at 408~K (Fig.~\ref{fig::ThermalTreatment}b). Sharpening is especially highlighted in averaged third order Bragg peaks ($\Gamma_3$). The measured superlattice peaks of oIC-CDW are sharper both in azimuthal (by $\sim$60\%) and radial (by $\sim$50\%) directions when compared to the IC-CDW. Notably, the superlattice peak widths of the oIC phase is comparable to the peak widths of the principle Bragg peaks. Therefore, the oIC is a spatially coherent electronic crystal.

The oIC-CDW, a 2D charge ordered state, is enhanced by at least one-hundred fold over previously reported bulk IC-CDWs.  Diffracted superlattice peaks in oIC-CDWs have an integrated intensity over ten times stronger despite that the number of charge ordered \tas{} layers has been reduced to less than 10\% of the material. Thus, endotaxial engineering improves not only the long range order but also the charge order amplitude of the IC-CDW. The correlation of long-range order and CDW enhancement is measured directly via hexatic CDW melting later in this manuscript. 

\section*{Endotaxial Polytype Heterostructure of \tas{}}
\begin{figure}
    \centering
    \includegraphics[width=0.95\linewidth]{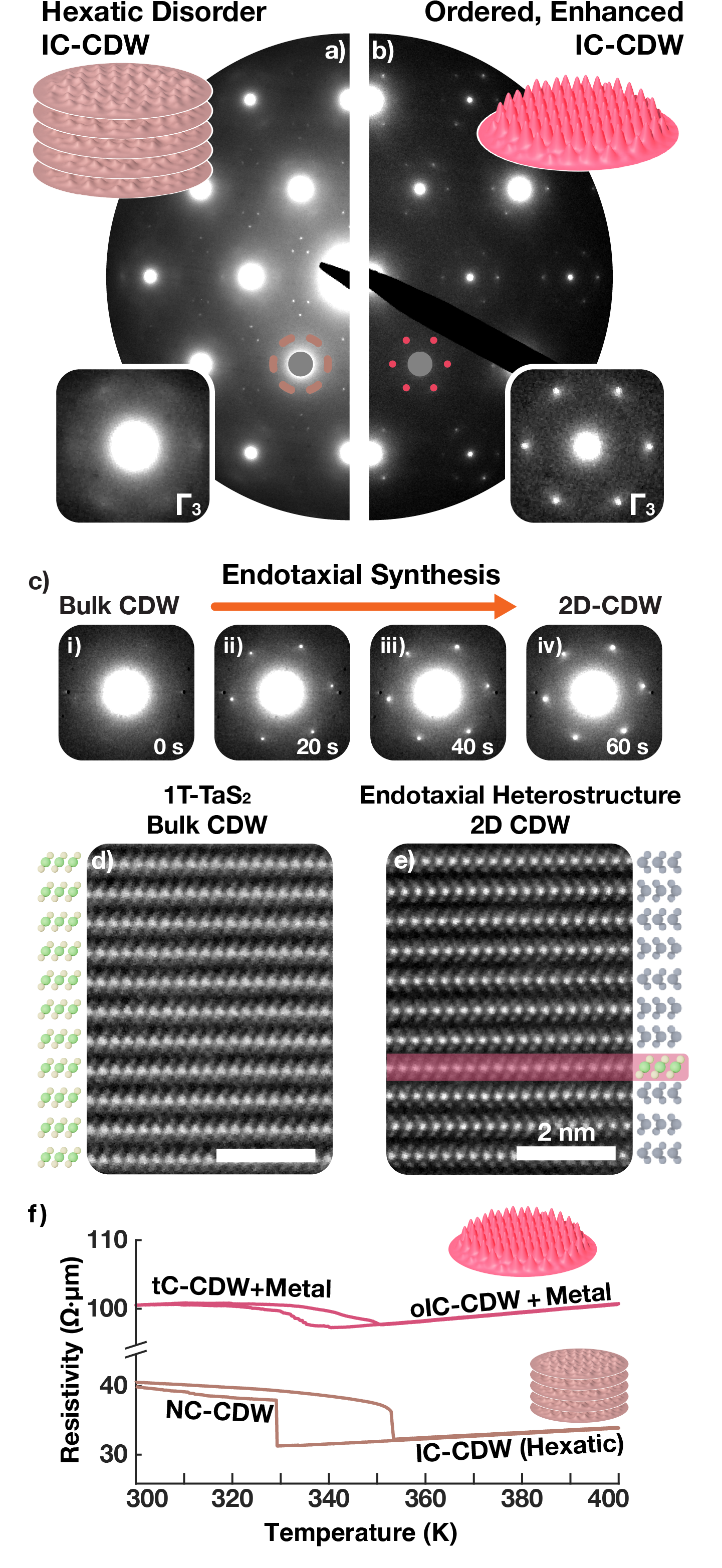}
    \caption{\textbf{Endotaxial polytype heterostructure of \tas{}}. a) In bulk \tas{}, an IC-CDW phase emerges above 350 K, with azimuthally diffused superlattice peaks characteristic of hexatic disorder. b) oIC-CDW in endotaxial polytype heterostructure has enhanced long-range order and amplitude. Superlattice peaks are well-defined, sharper and brighter. c) Evolution of IC-CDW during the endotaxial synthesis. Atomic resolution cross-sectional HAADF-STEM of d) bulk and e) heat-treated \tasse{} confirms polytypic transformation. After treatment, Pr layers encapsulate monolayers of Oc layers. Scale bar is 2~nm. A selenium doped sample was imaged to enhance chalcogen visibility. f) Resistivity vs temperature measurement of bulk (brown) and thermally-treated (red) \tas{} shows a marked increase in resistivity in IC-CDW phases. In pristine sample IC-CDW gives way to nearly commensurate (NC-) CDW around 350~K. In polytype heterostructure, twinned commensurate (tC-) CDW emerges at a similar temperature range.}
    \label{fig::ThermalTreatment}
\end{figure}

The oIC-CDW phase reported herein is stabilized by synthesizing endotaxial polytype heterostructures of \tas{}, where oIC-CDWs reside in monolayers of octahedrally coordinated (Oc-) \tas{} embedded within prismatic (Pr-) \tas{} matrix and one-to-one atomic registry (Fig.~\ref{fig::ThermalTreatment}e). Endotaxial polytype heterostructures are synthesized by heating 1T-\tas{} at $\sim$720~K for 15--30 min in an inert environment. Notably, 1T-\tas{} is metastable and goes through Oc-to-Pr endotaxial layer-by-layer polytype transformation upon heating ($\gtrsim$ 620~K). In-situ SAEDs (Fig.~\ref{fig::ThermalTreatment}c i--iv) were acquired at 20 seconds intervals at 408~K through the high temperature conversion process (723~K). These snapshots reveal sharpening of superlattice peaks---a clear indicator of enhanced CDW order. Cooling the sample midst transition stops the conversion and an interleaved polytype heterostructure is synthesized---confirmed by cross-sectional ADF-STEM.

Figure~\ref{fig::ThermalTreatment}d and e show atomic resolution micrographs of bulk 1T endotaxially converted to a polytype heterostructure. The atomic resolution images demonstrate endotaxial monolayer encapsulation of Oc-\tas{} (Fig.~\ref{fig::ThermalTreatment}e, highlighted red) in Pr-layers. The Pr-\tas{} (bulk: 2H, 3R) are metallic above $\sim$100~K. Previous work showed these metallic layers decouple CDWs out-of-plane and raise the critical temperature for commensurate quantum states (i.e., C-CDW) from $\sim$200~K to $\sim$350~K~\cite{Sung_2022_tCCDW}. 


Surprisingly, the endotaxial polytype heterostructure stabilizes long-range order in IC-CDWs at elevated ($\gtrsim$~350~K) temperatures. The oIC-CDW phase has correlation length comparable to the crystal lattice, quantified by comparing widths of both superlattice and Bragg peaks from in-situ selected area electron diffraction patterns (SA aperture: 850~nm diameter). This indicates the CDW is relatively ordered (i.e. spatially coherent) over the distances comparable to the parent atomic crystal ($\sim$10\textsuperscript{2}~nm).

This enhancement of long-range CDW order is accompanied by a marked increase of the in-plane resistivity of the IC phase (Fig.~\ref{fig::ThermalTreatment}f). Figure~\ref{fig::ThermalTreatment}f shows temperature vs in-plane resistivity measurement of 1T (brown) and endotaxial (red) specimen. Resistivity of endotaxial \tas{} is higher for IC-CDW phases ($>$358~K), despite having many metallic layers introduced to the system. This implies that oIC-CDWs have a much higher resistivity than hexatic-IC in 1T-\tas{}.

\begin{figure}
    \centering
    \includegraphics[width=1.0\linewidth]{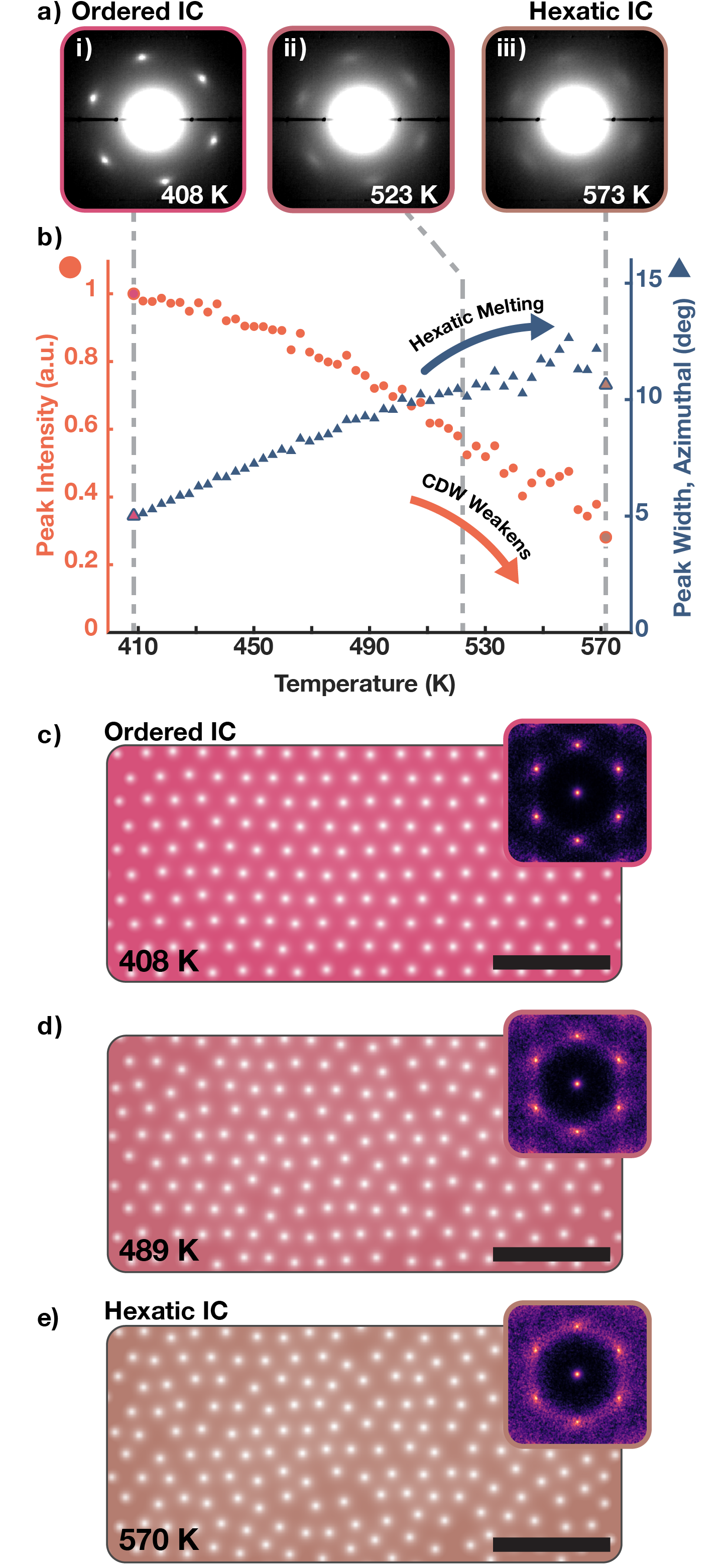}
    \caption{\textbf{Hexatic Melting of IC-CDWs.} a) Averaged in-situ SAED patterns showing oIC-CDW superlattice peaks in endotaxial heterostructure. i---iii) As temperature increases (408~K, 523~K, 573~K), superlattice peaks continuously blur along azimuthal direction. b) Quantification of superlattice peak profile. b-blue) Azimuthal width of the peak continuously increases with temperature---a key feature in hexatic melting process. b-red) Integrated superlattice peak intensity of oIC phase monotonically decays as temperature increases despite the increase in peak width; CDW is weakening. c--e) Monte Carlo simulation of 2D Lennard-Jones crystal with increasing temperatures. This represents the charge density distribution. As temperature increases, the crystal progressively disorders with increasing numbers of disclinations and dislocations. Insets) Structure factor of the simulated crystals. six-fold symmetry is apparent. As temperature increases, peaks diffuse prominently along azimuthal direction.}
    \label{fig::HexaticMelting}
\end{figure}

\section*{Hexatic Melting of IC-CDW}

Creating the oIC-CDW provides an ordered charge lattice that can be hexatically melted upon further heating. Hexatic melting is a uniquely 2D process wherein a crystal melts in two stages through the creation of dislocations and disclinations~\cite{Kosterlitz_1972,Kosterlitz_1973,Halperin_1978,Nelson_1982, Young_1979}. During this process the reciprocal space structure continuously evolves. Initially at lower-temperatures (c.a. 350 K), the oIC phase is an ordered charge crystal with well-defined peaks in reciprocal space (Fig.~\ref{fig::HexaticMelting}c). As temperature rises, the CDW peaks continuously blur azimuthally as the density of dislocations and disclinations increases (Fig.~\ref{fig::HexaticMelting}d, e). Azimuthal blurring of the reciprocal lattice is characteristic of hexatic phases and reflects the loss of translational symmetry while maintaining some orientational order~\cite{Brock_1989}. Eventually, at higher temperatures (c.a. 570 K), the hexatic crystal completely dissociates into an amorphous liquid state with ring-like structure factor. Figure~\ref{fig::HexaticMelting}c--e, are generated using a phenomological Monte Carlo simulation wherein displacement of the CDW charge centers follow a temperature dependent Maxwell-Boltzmann probability distribution (See Methods). Here, the incommensurate CDW hexatically melts while the underlying atomic lattice remains unchanged---in diffraction this corresponds to a blurring of CDW superlattice peaks and preservation of Bragg peaks. 

During the hexatic melting of oIC-CDWs, superlattice peaks increasingly blur as temperature is raised---clearly visible in in-situ SAED at Fig.~\ref{fig::HexaticMelting}a-i) 473~K, Fig.~\ref{fig::HexaticMelting}a-ii) 523~K, and Fig.~\ref{fig::HexaticMelting}a-iii) 573~K. The blurring is anisotropic and more prominent along azimuthal directions as expected for hexatic phases. The CDW peaks are quantified throughout the melting process in Figure \ref{fig::HexaticMelting}b. Azimuthal peak width (Fig.~\ref{fig::HexaticMelting}b, blue-triangles) increases continuously with temperature; roughly doubling when raised from 410~K to 570~K. Around 520~K the oIC has melted into a state that resembles the well-known IC-CDW for bulk \tas{}. This CDW melting process is reversible and peaks sharpen when temperature is decreased. Notably, Bragg peaks do not show appreciable changes indicating only the electronic crystal is melting, not the \tas{} atomic crystal.

Although the CDW melting process appears hexatic, it is distinct from familiar liquid crystals, silica spheres, or atomic crystals wherein the amplitude of the order parameter does not change. Here, quantitative analysis of the superlattice peak intensities (Fig.~\ref{fig::HexaticMelting}a-red) reveals the charge density wave amplitude decreases with temperature. This is expected as topological defects in CDWs (dislocations and disclinations) have locally divergent strain with elastic energy cost that forces a local amplitude collapse. These local CDW amplitude collapses have been observed at the center of topologcal defects in the 3D charge ordering of manganites~\cite{Savitzky_2017}.

\begin{figure*}[ht]
    \centering
    \includegraphics[width=1.0\linewidth]{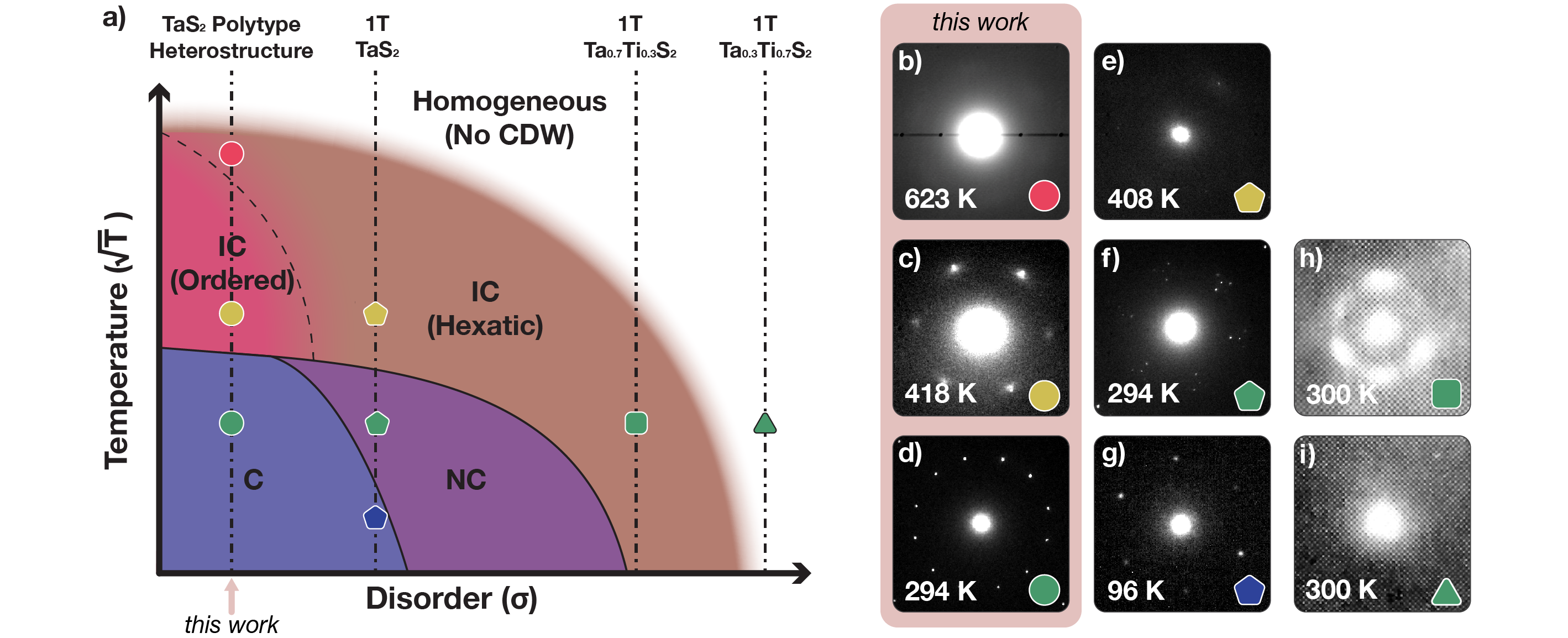}
    \caption{\textbf{Phase Diagram of Octahedrally Coordinated \tas{}.} a) Schematic temperature vs disorder phase diagram of octahedrally coordinated \tas{}. As extrinsic disorder ($\sigma$) decreases, more ordered CDW phases are stabilized. At room temperature, polytype heterostructures with low disorder stabilizes C-CDW (d) instead of NC-CDW (f), and long-range ordered IC-CDW (c) phase instead of hexatically disordered IC-CDW (e). Furthermore, it stabilizes CDWs (b) at higher temperatures then bulk 1T-\tas{} can (T\textsubscript{CDW} $\cong$ 540~K~\cite{Bayliss_1983}). Substitutional disorder, on the other hand, destroys long-range order and hexatic IC-CDW is stable at room temperature~(h) and leads to complete destruction of CDW eventually~(i). b--i) Electron diffraction patterns showing superlattice peaks around a single Bragg peak reveals the charge ordering states. h,i) are adapted from Wilson et al.~\cite{Wilson_1975}.}
    \label{fig::PhaseDiagram}
\end{figure*}

\section*{The CDW Phase Diagram for Octahedral \tas{}}

Endotaxial synthesis of octahedrally coordinated \tas{} allows access to new phases of matter and construction of a phase diagram for CDWs using temperature (T) and disorder (\textsigma{}). The CDW phase diagram for 1T-\tas{} is shown in Figure~\ref{fig::PhaseDiagram}. 1T-\tas{} exists with native disorder and the ordered, commensurate phase (C-CDW, Fig.~\ref{fig::PhaseDiagram}g) is only observed at low-temperatures. At room temperature, the CDW is a partially-ordered NC phase (Fig.~\ref{fig::PhaseDiagram}f) that enters the hexatic IC phase upon heating (Fig.~\ref{fig::PhaseDiagram}e). At high-temperatures or high-disorder, CDWs degrade or vanish. The high disorder regime was historically achieved by substituting tantalum ions with other metal species (e.g. Ti, Nb) or by forcing intercalates within the van der Waals gap~\cite{Wilson_1975}. At room temperature, mild substitution of titanium (1T-Ta\textsubscript{0.7}Ti\textsubscript{0.3}S\textsubscript{2}) drives the system into hexatic-IC CDW states (Fig.~\ref{fig::PhaseDiagram}h), and as more titanium is substituted (1T-Ta\textsubscript{0.3}Ti\textsubscript{0.7}S\textsubscript{2}) CDW vanishes completely (Fig.~\ref{fig::PhaseDiagram}i).

The low disorder regime, now accessible by endotaxial engineering, provides room temperature ordered C-CDWs and a novel ordered IC-CDW at higher temperatures. Notably with low-disorder, the C to IC transition is direct and the NC phase does not appear. The IC phase is ordered, but the CDW can be continuously melted into a disordered hexatic-IC phase (as described in figure~\ref{fig::HexaticMelting}). The boundaries of the CDW phase diagram are drawn with consistency to hexatic melting of 2D collidal particles under temperature and disorder~\cite{Deutschlander_2013} as well as nematic CDWs~\cite{Nie_2014, Yao_2022, Frachet_2022}.

Notably, CDWs in endotaxial \tas{} are two dimensional and the oIC phase has enhanced order despite the 3D to 2D dimensionality reduction. In bulk 1T-\tas{} CDWs are quasi-2D with non-negligible out-of-plane interaction~(Fig.~\ref{fig::Overview}h)~\cite{Naito_1984,Tanda_1984,Ritschel_2015,Stahl_2020}. Formation of endotaxial polytype heterostructures disrupts the out-of-plane interactions and CDWs reside in a protected 2D environment~\cite{Sung_2022_tCCDW}. Stabilization of an ordered IC-CDW in 2D seemingly contradicts with Hohenberg-Mermin-Wagner theorem~\cite{MerminWagner_1966, Hohenberg_1967} and Imry-Ma argument~\cite{Imry_1975} which state spontaneous symmetry breaking of continuous symmetry (e.g. IC-CDWs) is unstable at non-zero temperatures in 2D. While both principles do not prevent intermediate phases with short-range order, the 2D CDWs should be none-the-less more fragile to disorder~\cite{Nie_2014}. An ordered IC phase can only emerge in ultra-clean environments. Here endotaxial synthesis protects CDW states by strain-free encapsulation in a chemically identical environment of metallic layers that shield disorder.



\section*{Conclusion}

In summary, we demonstrate that endotaxial synthesis of clean interleaved polytypic heterostructures can stabilize fragile quantum phases such as ordered CDWs even at high temperatures. Here, we stabilize and enhance 2D charge density waves (both long-range order and amplitude) in an endotaxially confined monolayer of 1T-\tas{}. Surprisingly, the low-dimensional symmetry breaking of an ordered incommensurate CDW (oIC-CDW) appears, suggesting the quantum states reside within minimal extrinsic disorder. By enhancing CDW order the hexatic nature of IC-CDWs are revealed. Experimental observation matches advanced simulation of electron diffraction of charge lattices to provide the real-space evolution of 2D CDW melting. Heating the oIC-CDW in-situ TEM above 400~K we see a reversible hexatic melting process, in which disclinations and dislocations destroy long-range translational symmetry of the CDW while maintaining its orientational order.  The CDW melts well before the underlying atomic crystal changes. In 2D, CDWs are expected to manifest through vestigial electronic hexaticity---a weak CDW with substantial defects and short range order. The nature of vestigial phases in CDWs remains poorly understood with little direct evidence. From these results, a CDW phase diagram for 1T-\tas{} is created and consistent with the predicted emergence of vestigial quantum order.




\section*{References}
\par\vspace{5pt}
{\printbibliography[heading=none]}

\FloatBarrier

\section*{Acknowledgements}
S.H.S. acknowledges the financial support of the W.M. Keck Foundation. Experiments were conducted using the Michigan Center for Materials Characterization (MC2) with assistance from Tao Ma and Bobby Kerns. This work made us of electron microscopy facility of the Platform for the Accelerated Realization, Analysis, and Discovery of Interface Materials (PARADIM) supported by the National Science Foundation, which is supported by National Science Foundation under Cooperative Agreement No.~DMR-2039380. N.S. acknowledges additional support from the NSF GRFP under award number DGE-2139899. P.K. and J.H. gratefully acknowledge support from NSF MRSEC DMR-2011839. Y.L, W.J.L. and Y.P.S, thank the support from the National Key
R\&D Program (Grant No.~022YFA1403203 and No.~2021YFA1600201), the National
Natural Science Foundation of China (Grant No.~U2032215, No.~U1932217 and
No.~12274412).

\section*{Author contributions}
S.H.S and R.H. conceived the charge lattice model and associated lattice distortions and linked them to diffraction of \tas{}. S.H.S., Y.M.G., N.S., L.F.K., and  R.H. performed HAADF-STEM and in-situ TEM and interpreted electron microscopy data. S.H.S. fabricated samples for electronic measurements. P.K. and J.T.H. performed and analyzed electronic measurements. S.H.S., I.E.B., R.H. and K.S. provided theoretical interpretation. S.H.S. and N.A. performed Monte-Carlo simulations. S.H.S. K.S. and R.H. created the phase diagram of octahedrally coordinated \tas{}. Y.P.S. synthesized 1T-\tasse{} crystal. S.H.S. and R.H. prepared the manuscript. All authors reviewed and edited the manuscript.

	\section*{Competing Interests}
The authors declare no competing interests.

\section*{Methods}
\subsection*{Simulated Diffraction of Charge Lattices with Heating}
Charge density waves are electronic modulations describable in reciprocal space by three wave vectors (so called, triple q) or in real-space as local charges arranged into a hexagonal lattice. For a fully ordered system, the charge lattice is a perfect lattice (Fig.~\ref{fig::Overview}b left), and the structure factor (Fig.~\ref{fig::Overview}b left inset) is also a perfect lattice.  Here, the periodicity is equal to the incommensurate CDW wave vector \textbf{q}\textsubscript{IC} (or \textbf{a}\textsubscript{IC} in real-space). Traditional CDW theory elegantly describes ordered (or slightly disordered) systems using sparse representation in reciprocal space for ordered systems. However, a real-space basis readily describes topological disorder (dislocations and disclinations) in a charge density wave. This becomes particularly critical for IC phase ($>$350~K) of 1T-\tas{}, where diffraction studies reveal azimuthally diffused superlattice peaks~\cite{Wilson_1975} that we show to be consistent with topological disorder in CDWs. Describing disorder of CDW plays a critical role in simulating experimentally consistent diffraction patterns at high temperatures.

The hexatic melting of a real-space charge lattice is illustrated with phenomenological Monte Carlo simulations of the NPT ensemble (constant particle count, temperature, and pressure). The displacement of charge centers in a CDW follow a Maxwell-Boltzmann probability distribution at different temperatures. The interaction energy between charge centers is calculated using a shifted Lennard Jones potential truncated at 18.7 \AA. From these first principles, the likelihood of forming dislocations and disclinations in a CDW lattice increases with temperature.

Diffraction of the simulated CDWs is calculated from the corresponding periodic lattice distortion (PLD) of a 1T-\tas{} crystal. The displacements are small ($\lesssim$10 pm), but clearly manifest as superlattice peaks with distinctive intensity in SAED. Notably, the superlattice peak intensities becomes stronger at higher $|\mathbf{k}|$; this is distinguishable from chemically ordered superlattice peaks that decay as $|\mathbf{k}|$ increases~\cite{Sung_2022_tPLD}. In \tas{}, atoms displace toward the charge centers which is equivalent to a longitudinal displacement wave. Here, a the displacement amplitude is proportional to the charge density gradient with a max displacement set at 7 pm. Electron diffraction is kinematically simulated under a flat Ewald Sphere approximations using the Fourier transform of the displaced atomic lattice.

\subsection*{Electron Microscopy}
In-situ SAED was performed on Thermofisher Scientific (TFS) Talos (operated at 200 keV, SA aperture 850~nm) with Protochips Fusion Select holder and Gatan OneView Camera. Cross-sectional HAADF-STEM images were taken on JEOL 3100R05 (300~keV, 22~mrad) with samples prepared on TFS Nova Nanolab DualBeam FIB/SEM. 

TEM specimens were prepared by exfoliating bulk 1T-\tas{}{} and 1T-\tasse{} crystals onto polydimethylsiloxane (PDMS) gel stamp. The sample was then transferred to TEM grids using home-built transfer stage. Silicon nitride membrane window TEM grid with 2~\textmu{}m holes from Norcada and Porotochips Fusion Thermal E-chips. From optical contrast and CBED patterns, the samples (Fig.~1, 2) were estimated to be 20--50~nm thick~\cite{Li_2013, Hovden_2018}.

\subsection*{Synthesis and Acquisition of bulk crystals}
1T-\tas{} for in-situ SAED measurements and electronic measurements was acquired from HQ Graphene. 1T-\tasse{} (x~$\approx$~1) for cross-sectional HAADF-STEM measurements was grown by the chemical vapor transport method with iodine as a transport agent. Stoichiometric amounts of the raw materials, high-purity elements Ta, S, and Se, were mixed and heated at 1170~K for 4 days in an evacuated quartz tube. Then the obtained \tasse{} powders and iodine (density: 5 mg/cm\textsuperscript{3}) were sealed in another longer quartz tube, and heated for 10 days in a two-zone furnace, where the temperature of source zone and growth zone was fixed at 1220~K and 1120~K, respectively. A shiny mirror-like sample surface was obtained, confirming their high quality. All CDW characterization was done on 1T-\tas{}; Se-doped sample was used only for polytype characterization in cross-sectional HAADF-STEM (Fig.~\ref{fig::ThermalTreatment}d,e).

\subsection*{Endotaxial Synthesis of oIC-CDW in \tas{}}
Interleaved 2D \tas{} polytypes were synthesized by heating 1T-\tas{} to 720 K in high vacuum ($<$10\textsuperscript{\textminus{}7} Torr) or in an argon purged glovebox~\cite{Sung_2022_tCCDW}. 1T-\tas{} was held at 720 K for $\sim${10} minutes, then brought down to room temperature. Once the interleaved polytype is fully established, the oIC-CDW becomes stable electronic state above 350~K.

\subsection*{Device Fabrication and Electronic Measurement}
For resistivity measurements, \tas{} flakes were transferred using PDMS gel stamp method to pre-fabricated bottom contacts. The fabrication of bottom contacts are detailed in~\cite{Sung_2022_tCCDW}. The flake was sculpted into rectangular bar ($\sim$11~\textmu{}m$\times$15~\textmu{}m) using TFS Nova Nanolab DualBeam FIB/SEM (See Supplementary Figure~S4). The thickness of the flake was determined by AFM.

Resistivity vs temperature measurements were performed in a Quantum Design Dynacool PPMS using a standard sample puck and an external Keithley 2400 series source meter. The sample was adhered to the puck backplane with silver paint, and contacts were wire bonded to the puck channel pads using 50~\textmu{}m Au wire. To ensure sample thermalization, a baffle rod with an Au-coated sealing disk hovering $<$1~cm above the sample was inserted into the PPMS bore, and the heating and cooling rate was restricted to $<$2~K/min. 10~\textmu{}A current was sourced for four wire measurements. The current/voltage limits were chosen to keep electric fields below 10~kV/cm to avoid sample breakdown, as well as to keep current densities below 10\textsuperscript{5}~A/cm\textsuperscript{2} and prevent localized heating at low temperatures.

\end{document}